\documentstyle[11pt]{article}
\def\la{\mathrel{\mathchoice {\vcenter{\offinterlineskip\halign{\hfil
$\displaystyle##$\hfil\cr<\cr\sim\cr}}}
{\vcenter{\offinterlineskip\halign{\hfil$\textstyle##$\hfil\cr<\cr\sim\cr}}}
{\vcenter{\offinterlineskip\halign{\hfil$\scriptstyle##$\hfil\cr<\cr\sim\cr}}}
{\vcenter{\offinterlineskip\halign{\hfil$\scriptscriptstyle##$\hfil\cr<\cr\sim
\cr}}}}}

\newcommand{\be}{\begin{equation}}
\newcommand{\ee}{\end{equation}}
\newcommand{\bea}{\begin{eqnarray}}
\newcommand{\eea}{\end{eqnarray}}

\def\l{\left}
\def\r{\right}
\def\dis{\displaystyle}
\begin{document}
\begin{center}
{\bf Dark Matter -- Possible Candidates and Direct Detection}
\end{center}
\begin{center}
Debasish Majumdar \\
Saha Institute of Nuclear Physics, \\
1/AF Bidhannagar, Kolkata 700 064, India
\end{center}

\begin{center}
{\bf Abstract}
\end{center}
{\small 
The cosmological observations coupled with theoretical calculations suggest 
the existence of enormous amount of unseen and unknown matter or 
dark matter in the universe. The evidence of their existence, the possible
candidates and their possible direct detections are discussed. }  

\section{Introduction}
The observations by Wilkinson Microwave Anisotropy Probe or 
WMAP \cite{wmap} for 
studying the fluctuations in cosmic microwave background radiation reveal 
that the universe contains 27\% matter and the rest 73\% is an unknown energy 
known as Dark Energy. Out of this 27\%,
only 4\% accounts for the ordinary matter like leptons and baryons, 
stars and galaxies etc. The rest 23\% is completely unknowm. 
Moreover, there are strong indirect evidence (gravitational) from various 
observations
like velocity curves of spiral galaxies, gravitational lensing etc. in
favour of the existence of enormous amount of invisible, nonluminous matter
in the universe.
The measurement of mass-luminousity ratio which can be used 
to determine the cosmological density parameter also estimates 
a very low value for luminous matter. 
This huge amount of unknown and ``unseen" matter (which in fact 
constitutes more than 90\% of the total matter content of the universe) is   
known as ``Dark Matter". 

Although the nature and identity of dark matter still remain 
a mystery, indirect evidence suggests that they are stable and probably heavy,
non relativistic (Cold Dark Matter or CDM) and are weakly
interacting. Therefore they are often known as Weakly Interacting Massive
Particles or WIMPs.

In this article, the properties, types and the possible candidates 
of dark matter are discussed. The possibilities
of their direct detection and theoretical detection rates are also 
given.

\section{Cosmological Density Parameter}

The space-time metric consistent with the homogeneity and isotropy 
of the universe -- on large scales -- can be given by the 
Robertson-Walker (RW) metric 
\begin{equation}
ds^2 = -dt^2 + a^2(t) \left [ \frac {dr^2} {1 - kr^2} 
                              + r^2(d\theta^2 + \sin^2\theta d\phi^2) \right ]
\end{equation}
Here $a(t)$ is a scale factor and $k$ denotes the spatial curvature. 
Thus, $k=+1$ means the spatial section is positively curved, i.e. 
the space is locally isometric to 3-D spheres; $k=-1$ signifies
that the space is locally hyperbolic (spatial section is negatively curved);
and finally $k=0$ signifies no spatial curvature, i.e. a flat geometry 
for the local space. 

The RW metric follows from the kinematic consequences. The dynamics, i.e. the 
time evolution of the scale factor $a(t)$ follows by applying Einstein's 
equation (with the cosmological constant $\Lambda$)
\begin{equation}
R_{\mu\nu} - \frac {1} {2} g_{\mu\nu} R = 8\pi G T_{\mu\nu} + \Lambda
\end{equation}
to the RW metric. 
In the above, $R_{\mu\nu}$ are Ricci Tensors, $R$ is the Ricci scalar,
$g_{\mu\nu}$ is the spatial
metric, $T_{\mu\nu} = (\rho + p) U_\mu U_\nu + pg_{\mu\nu}$ is the 
energy-momentum tensor contains the density
$\rho$ and pressure $p$. The Einstein's 
equation relates the geometry with the energy-momentum. 

Applying Einstein's equation to cosmology, one gets the Friedmann's 
equation
\begin{equation}
\left ( \frac {1} {a} \frac {da} {dt} \right )^2 = \frac {8 \pi G} {3} \rho
              + \frac {\Lambda} {3}    - \frac {k} {a^2}
\end{equation}  
Defining $\frac {1} {a} \frac {da} {dt} = H$, the expansion rate of 
the universe or formally Hubble constant, the above equation can be 
written as 
\begin{equation}
\frac {k} {H^2a^2} = \frac {8\pi G} {3H^2} \rho + \frac {\Lambda} {3 H^2} - 1
\end{equation}
Defining $\frac {3H^2} {8\pi G} = \rho_c$ -- the critical density of 
the universe, the above equation takes the form 
\begin{eqnarray}
\frac {k} {H^2a^2} &=& \frac {\rho} {\rho_c} + \frac {\rho_{\Lambda}} {\rho_c} 
                     - 1 \nonumber \\
&=& \Omega_m + \Omega_\Lambda - 1 
\end{eqnarray}
where $\Omega_m = \frac {\rho} {\rho_c} $ and  
$\Omega_\Lambda = \frac {\rho_{\Lambda}} {\rho_c} $ are the 
cosmological density parameters for matter and energy respectively.
For a flat universe ($k =0$) we have therefore
\begin{equation}
\Omega = \Omega_m + \Omega_\Lambda = 1
\end{equation}

The analysis of WMAP probe predicts curvature parameter $k=0$ (the 
universe is spatially flat)\footnote{A stringent limit is however put for
$\frac {k} {H^2 a^2} = \Omega_k = -0.003 \pm 0.010$ \cite{chris}.} 
and therefore the matter density of the 
universe 
$$
\Omega_m = \Omega_{\rm visible} + \Omega_{\rm DM} = 0.27
$$
out of which 
$$
\Omega_{\rm visible} = 0.4\,\,\,\,\,\,\,\,\,
{\rm and}\,\,\,\,\,\, \Omega_{\rm DM} = 0.23
$$
where `DM' stands for the dark matter and the energy 
density (unknown dark energy) $\Omega_\Lambda$ is
$$
\Omega_\Lambda = 0.73 
$$

\section{Evidence of the existence of Dark Matter}

The evidence of dark matter was first envisaged by the observation of 
motion of galaxies in cluster of galaxies like Virgo and Coma. A galaxy
cluster is a gravitationally-bound group of galaxies\footnote{A cluster 
can be rich with thousand(s) of galaxies or can be poor with $\sim$ 
30 - 40 galaxies. The cluster, Local Group, to which our galaxy $-$ Milky Way 
$-$ belongs contains only about 30 galaxies.}. Assuming the dynamical 
equilibrium of the cluster, it obeys the Virial theorem, 
$K + U/2 = 0$, where $K$ is the kinetic term and $U$ the potential. 
The kinetic term $K$ was estimated by measuring the velocities 
of individual galaxies and is found to be much larger than the 
potential term $U$ which was calculated by assuming that the mass of 
the cluster is the sum of the individual mass of the galaxies. 
This discrepancy indicates the existence of unseen and unknown mass
in the cluster. 

Stronger observational evidence exists by studying the rotational 
velocities of the stars inside a galaxy (rather than observing the 
galaxy itself inside a cluster). For a star in a spiral galaxy 
$-$ which can be 
considered as a rotating disc with a central bulge where most of 
the galactic mass is concentrated $-$ describing a circular orbit 
at a radial distance $r$ from the centre of the galaxy, 
with a rotational velocity $v_r$, one has
\begin{equation}
\frac {mv_r^2} {r} = \frac {G M_r m}{r^2}
\end{equation}
where $m$ is the mass of the star and $M_r$ is the mass inside the 
orbit of radius $r$. If the object is inside the central concentrated mass 
region, the mass $M_r$ can be estimated as 
\begin{equation}
M_r = \frac {4} {3} \pi r^3 \rho 
\end{equation}
where $\rho$ is the average density of the central region. From 
Eq. (7) therefore we readily see 
\begin{equation}
v_r \sim r 
\end{equation}
Now, for a star outside the central bulge, one can approximate 
$M_r = M$ (a constant, neglecting the mass outside the central 
bulge) and in this case the nature of rotational velocity $v_r$ 
becomes (from Eq. (7))
\begin{equation}
v_r \sim \frac {1} {r^{1/2}}\,\,\,\,\, ({\rm Keplerian}\,\,\,{\rm Decline})
\end{equation}
Hence normally, for rotational velocities $v_r (r)$, one would expect an 
initial rise with increase of radial distance $r$ from the 
galactic centre (Eq. (7)) and then a Keplerian decline for radial 
distance $r$ outside the central bulge. 

Instead, the observation of rotation curves (variation of $v_r$ with $r$)
reveal the initial rise of $v_r$ with $r$ as expected but then 
$v_r(r)$ becomes a constant with the increase of $r$ instead of suffering 
the $r^{-1/2}$ decline. Hence from Eq. (7), with $v_r$ constant 
\begin{equation}
M_r \sim r
\end{equation}
which suggests the existence of enormous unknown mass. 
     
The evidence of dark matter is indicated from other observations 
like measurement of temperature and density of hot X-ray emitting 
gases from elliptical galaxies like M87.

The other evidence comes from the observance of the phenomenon of 
gravitational lensing. This occurs due to the bending of light 
in presence of gravitational potential. The mass of a cluster 
and hence $\Omega_m$ can be estimated by exploring the multiple 
lense effects of  background galaxies produced by the cluster. 
These observations also point to  $\Omega_m \sim 0.3$.  

\section{Types of Dark Matter}

On the basis of the nature of the constituents, the dark matter can be divided 
into two types namely a) baryonic and b) non-baryonic. Different cosmic 
microwave anisotrpy (CMB) measurements predict a value of baryon density  
to be $\Omega_b \sim 0.04$ which is far less then the total dark matter
density $\Omega_{\rm DM} = 0.23$. This is indicative of the fact that 
the most of the dark matter in the universe is non-baryonic in nature. 

Again, on the basis of their velocities, the dark matter can be 
broadly classified as a) Hot Dark Matter (HDM) and b) Cold Dark Matter (CDM). 
For HDM, the particle candidates are light and hence move with 
relativistic velocity while the CDM candidates are heavy and move 
with non-relativistic velocities. If a candidate falls in between 
the two categories they are sometimes referred to as Warm Dark Matter. 

Neutrinos can be a possible candidate for Hot Dark Matter, but their 
relic density falls far short of the total dark matter density, 0.23, 
if the neutrinos are indeed light ($\sim$ eV).  
It is general wisdom that, most of the dark matter of the 
universe is Cold type (CDM) and non-baryaonic in nature.     

\section{Candidates for Dark Matter}

The dark matter candidates still remain an enigma. 
But the fact that they constitute more than 90\% of the 
matter content of the universe and their little or no interaction
with any Standard Model particles indicate that they are made up 
of stable, neutral and very weakly (or almost non-) intearcting  particles.
Also most of their constituents are massive (heavy) to account 
for that large mass. 
 
The known particles like baryons are proposed but as is discussed 
earlier baryons alone cannot explain the total dark matter of the 
universe. But some of the dark matter may need to be baryonic 
as $\Omega \la 0.01$ in the galactic disk. 

There are other candidates (baryonic) for dark matter like 
jupiter-like objects, dead massive stars etc. But they fail to
account for the density $\Omega_{\rm DM} = 0.23$.  

Recently there has been experimental evidence of at least one form of
dark matter namely Massive Astrophysical Compact Halo Objects or
MACHOs in the halo of Milky Way galaxy. The light from a distant star,
passing by a MACHO, bends due to the large gravitational field of the
MACHO. The bending of light is a consequence of Einstein's General
Theory of Relativity and as discussed above, is known as gravitational lensing.
In the present case, since the lens is relatively small (compared to galaxy),
multiple images are not observed. On the other hand, due to relative
motion between the stars and MACHOs the lensing effect causes an increase
in the brightness of that distant object. Using this phenomenon,
known as gravitational microlensing, around 13-17 MACHOs have been
detected in the Milky Way Halo.
                                                                                
A candidate for MACHOs has been proposed in Ref. \cite{sinha}. 
It is suggested that MACHOs have evolved
out of the strange quark nuggets (SQNs) formed during the first order
phase transition of the early universe from quark phase to hadronic phase
at a temperature around 100 MeV ($\sim 10^{-5}$ second after Big Bang).
During this phase transition, hadronic matter starts to appear as
individual bubbles in quark-gluon phase \cite{Witten,ISO}. With the
progress of time more bubbles appear and they expand to form a network
of such bubbles (percolation) in which the quark matter gets trapped.
With further cooling of the universe, these trapped domains of quark
matter shrink very rapidly without significant change of baryon number
and eventually evolves to SQNs through weak interactions with almost
nuclear density \cite{raha}. These objects are stable and calculation
shows that to explain all the CDM, the baryon number of an SQN should
be $\sim 10^{42-44}$ \cite{bhatta} assuming all SQNs to be of same size.
These SQNs with masses $\sim 10^{44}$ GeV and size $\sim 1$ metre,
would have very small kinetic energy compared to their mutual Gravitational
potential.

Among the possible candidates of light non baryonic dark matter, 
come the relic neutrinos. But as briefly discussed earlier, the 
light neutrinos cannot account for the dark matter relic density 
obtained from, say, WMAP observation. 

Another viable light dark matter candidate is axion. Axion is a 
pseudo-Goldstone boson and is introduced to solve the strong CP 
problem \cite{axion} (conservation of CP symmetry in Quantum chromodynamics
or QCD). It arises as a consequence of a global 
$U(1)$ symmetry (Peccei-Quinn symmetry). The axions gets a 
small mass due to the breaking of this global $U(1)$ symmetry. 
Axions can also be produced in supernova. But the QCD consideration 
alongwith the production process of axions in supernova \cite{axionsn}
(through nucleon-nucleon Bremsstrahlung), it is estimated that axion can 
be a dark matter candidate within a very limited window \cite{olive}.

For the particle candidates of Cold Dark Matter or CDM that are 
non-baryonic in nature, there are various proposals. These candidates 
are not Standard Model (SM) particles and follow from the theories 
beyond SM like Supersymmetric theories or theories with extra dimensions. 
These particles if existed would have manifested themselves 
at higher energy scales during the very early phase of the universe.
With the expansion of the universe, when the annihilation rate 
of these particles fall below the expansion rate 
of the universe, these particles get decoupled from the 
universe fluid and remain as they were. This phenomenon is 
known as ``freeze out". After the freeze out takes place those 
particles float around as relics.     

The popular and favourite candidate for non-baryonic CDM is proposed
from theory of Supersymmetry or SUSY. Supersymmetry is the symmetry between 
fermions and bosons or rather more precisely the symmetry between 
the fermionic and bosonic degrees of freedom. This is introduced to 
address the so called ``hierarchy problem" or ``Weak scale instability 
problem". The hierarchy such as W-boson mass $m_W << M_p$ or the SM 
Higgs Boson mass $m_H << M_p$, where $M_p$ is the Planck Mass 
($1/\sqrt{G_N} \sim 10^{19}$ GeV) tends to be destroyed as a 
consequence of the higher order correction to the mass.  
The correction suffers a quadratic divergence. A fine tune of large orders of 
magnitude is required to restore the physical SM Higgs mass. This fine
tuning in turn affects the masses of other SM fermions and gauge 
bosons and thus hierarchy. SUSY stabilises this hierarchy and peeps 
to the possibility of new physics beyond the electroweak energy scale
of $\sim 250$ GeV. 

In minimal supersymmetric standard model or MSSM (see e.g. \cite{drees}), 
each SM fermion
has their bosonic SUSY partner and the gauge bosons have 
their fermionic SUSY partners. Thus in the MSSM framework, 
one generation in SM is to be represented by 
five left handed chiral superfields $Q$, $U^c$, $D^c$, $L$, $E^c$
where the superfield $Q$ contains quarks and their bosonic superpartner,
squark $SU(2)$ doublets; $U^c$ and $D^c$ are the quark and squark 
singlets; $L$ contains leptons and their bosonic superpartner slepton 
$SU(2)$ doublets and $E^c$ contains lepton and slepton singlets. 
In the gauge sector however, in MSSM framework, in addition to 
the SM gauge bosons, we have eight 
gluinos, the fermionic superpartners of QCD gluons; three winos ($\tilde{W}$)
the fermionic partner of $SU(2)$ gauge bosons and a bino ($\tilde{B}$), 
the fermionic partner of $U(1)_Y$ gauge boson. 
In the Higgs sector, one needs to introduce two Higgs superpartners 
$\tilde {H_1}$ and $\tilde {H_2}$ in order to break the 
$SU(2) \times U(1)_Y$. Without going into
details, due to space constraints it is only mentioned that the 
two Higgsino doublet with hypercharge $Y = +1/2$ and $Y = -1/2$ 
make the model anomaly free (cancellation due to opposite hypercharge).

It is a general practice in MSSM (to ensure protection against 
rapid proton decay),
to introduce a parity called $R$ parity and  
it is assumed to be conserved.
The $R$ parity is defined as $R = (-1)^{3B + L + 2S}$, where, 
$B$ is the baryon number, $L$, the lepton number and $S$ the spin. 
This ensures that the Lightest Supersymmetric Particle or LSP is stable 
and if it is neutral then can be a candidate for dark matter. 

One such dark matter candidate is neutralino ($\chi$) \cite{jungman} which 
is the linear superposition 
of the fermionic superpartners of neutral SM gauge bosons and Higgs bosons 
and can be written as  
\begin{equation}
\chi = \alpha {\tilde{B}} + \beta {\tilde{W}}^0 + \gamma {\tilde {H}}_1 
           + \delta {\tilde {H}}_2
\end{equation} 
The coefficents can be obtained by diagonalizing the mass matrix 
(in the basis \{ ${\tilde{B}}$, ${\tilde{W}}^0$, ${\tilde {H}}_1$,
           ${\tilde {H}}_2$ \})

\begin{equation}
%
%
\left ( \begin{array}{cccc} 
M_2 & 0 & -M_Z \cos\beta\sin\theta_W & M_Z\sin\beta\sin\theta_W \\
0 & M_1 & M_Z\cos\beta\cos\theta_W & -M_Z\sin\beta\cos\theta_W \\
-M_Z \cos\beta\sin\theta_W & M_Z\cos\beta\cos\theta_W  & 0 & -\mu \\
M_Z \sin\beta\sin\theta_W & M_Z\sin\beta\cos\theta_W  & -\mu & 0 
\end{array} \right )\, .
%
\end{equation}
In the above, the parameters $M_1$ and $M_2$ are soft SUSY 
breaking terms, $\mu$ 
is the so called ``$\mu$ term" in the superpotential (associated with two Higgs
supermultiplates), $\tan\beta = \frac {v_2} {v_1}$, the ratio of the  
vev's of two Higgs. 

The lightest neutralino eigenstate (LSP) of the mass matrix above (Eq. (13)) 
is considered to be a candidate for dark matter.

Another important proposal for dark matter candidates comes from 
the theories of extra higher dimensions. Although we live in a four dimensional 
world, there is apparently no reason to believe that extra dimensions 
do not exist. If dimensions $>4$ do at all exist they must be so compactified
that the effect due to them is not manifested in our 4-D world.
The ideas and theories of extra dimensions have been proposed to 
look for new physics beyond standard model and to address the 
hierarchy problem mentioned earlier as also to 
explain the non SM particles like gravitons (unification of gravity
and gauge interactions), cosmological constant problem etc.   
 
The effect of compactification of one extra space dimension can be 
demonstrated by considering a Lagrangian density ${\cal L}$ for a 
massless 5 dimensional scalar field $\Phi$, where one extra spatial 
dimension is inculded \cite{gaba}. Thus (following \cite{gaba}) 
\begin{eqnarray}
\Phi \equiv \Phi (x_\mu, y), && \mu = 0,1,2,3\,;\,\,\, 
y\,\,{\rm is}\,\,{\rm the}
\,\,\,{\rm extra}\,\,\,{\rm spatial}\,\,\,{\rm coordinate} \nonumber \\   
{\cal L} &=& -\frac {1} {2} \partial_A \Phi \partial^A \Phi
\,\,\,\, A = 0,1,2,3,4 
\end{eqnarray}
The extra 5th dimension is compactified over a circle of radius $R$ 
so that at distance scales $>> R$, the radius of compactification, 
the effect of extra dimension 
is not manifested. It is to be noted that the field is periodic 
in $y \rightarrow y + 2\pi R$ ($\Phi(x,y) = \Phi(x,y+2\pi R)$).  
Thus, expanding $\Phi (x,y)$ in $y$ as 
\begin{equation}
\Phi(x,y) = \displaystyle \sum_{n = -\infty}^{\infty} 
\phi_n (x) e^{iny/R}
\end{equation}
(with $\phi_n^* (x) = \phi_{-n} (x)$)
and substituting in the expression for ${\cal L}$ in Eq. (14) we have 
\begin{equation}
{\cal L} = \displaystyle \frac {1} {2} \displaystyle 
\sum_{n,m = -\infty}^{\infty} 
\left ( \partial_\mu \phi_n \partial^\mu \phi_m +
\frac {nm} {R^2} \phi_n \phi_m \right ) e^{i(n + m)y/R}
\end{equation}
The action $S$ is given by 
\begin{eqnarray}
S &=& \int d^4 x \displaystyle\int_0^{2\pi R} dy \,\,\, {\cal L} 
\end{eqnarray}
Replacing ${\cal L}$ (using Eq. (16)) and integrating out the 5th dimension to 
obtain the equivalent four dimensional result, the action $S$ becomes
\begin{equation}
S = \displaystyle\int d^4 x \left (- \displaystyle\frac {1} {2} 
\partial_\mu \psi_0 \partial^\mu \psi_0 \right ) -
\displaystyle\int d^4 x \displaystyle\sum_{k=1}^\infty 
\left ( \partial_\mu \psi_k \partial^\mu \psi_k^* + \displaystyle\frac
{k^2} {R^2} \psi_k \psi_k^* \right )
\end{equation} 
where $\psi_n = \sqrt{2\pi R} \phi_n$. Thus, from Eq. (18) we see 
that for a massless scalar field in 5-dimension, compactification 
over a circle yields, in equivalent 4-dimensional
theory, a zero mode ($\psi_0$) as real scalar field and an 
infinite number (tower) of massive
complex scalar fields with tree level masses given by  $m_k = k/R$. These 
modes are known as Kaluza-Klein modes (or Kaluza-Klein tower) and the integer
$k$ becomes a quantum number called 
Kaluza-Klein (KK) number which corresponds to the quantized momentum
$p_5$ in the compactified dimension. The 5-D Lorentz invariance (local) 
of the tree level Lagrangian allows us to write the dispersion relation 
as 
\begin{equation}
E^2 = {\bf p}^2 + p_5^2 = {\bf p}^2 + m_k^2
\end{equation}
where {\bf p} is the usual 3-D momentum. 
The conservation of this KK number apparently
seems to indicate that the Lightest Kaluza-Klein Particle or LKP is stable 
and can be a possible candidate for dark matter. 

An LKP dark matter candidate is proposed by Cheng et al \cite{cheng}      
in the model of universal extra dimension (UED) \cite{appelquist,matchev}. 
According to UED model the extra dimension is accessible to all standard 
model fields. In other words all SM particles can 
propagate into the extra dimensional space. Therefore every SM 
particle has a KK tower. The proposed LKP 
candidate for dark matter in UED model is the first KK partner 
$B^1$, of the hypercharge gauge boson.    

But in order to obtain chiral fermions in equivalent 4-D theory,
the compactification over a circle ($S^1$) does not suffice. The 
simplest possibility for the purpose is to compactify the extra dimension 
over an orbifold $S^1/Z_2$ \cite{matchev2} where $S^1$ is the circle 
of compactification radius
$R$ and $Z_2$ is the reflection symmetry under which the 5th coordinate
$y \rightarrow -y$. The fields can be even or odd under $Z_2$ symmetry.  
This orbifold can be looked as a line segment of length
$\pi R$ such that $0 \leq y \leq \pi R$ with the orbifold 
fixed points (boundary points) at 
0, $\pi R$ with two boundry conditions (Neumann and Dirichlet) 
for even and odd fields given by,
\begin{eqnarray}
\partial_5 \phi &=& 0\,\,{\rm For}\,\, {\rm even}\,\, {\rm fields} \nonumber \\
\phi &=& 0\,\,{\rm For}\,\, {\rm odd}\,\, {\rm fields} 
\end{eqnarray}
A consistent assignment for chiral fermion $\psi$ would be; 
($\psi_L$ even, $\psi_R$ odd) or vice versa, for gauge field $A$; 
$A_\mu$ even ($\mu =0,1,2,3$), $A_5$ odd and the scalars can be 
either even or odd.

Now from Eq. (15) and using the orbifold compactification discussed 
above, the KK decomposition of $\Phi$ in even or odd fields looks as
\begin{eqnarray}
\Phi_+ (x,y) &=& \sqrt {\frac {1} {\pi R}} \phi_+^0 + \sqrt {\frac {2} {\pi R}}
\sum_{n=1}^{\infty} \cos \frac {ny} {R} \phi^n_+ (x) \nonumber \\
\Phi_- (x,y) &=& \sqrt {\frac {2} {\pi R}}
\sum_{n=1}^{\infty} \sin \frac {ny} {R} \phi^n_- (x) \,\,\, .
\end{eqnarray}
Thus, $\Phi_-$ (odd field) lacks a zero mode due to the effect of 
$Z_2$ symmetry and Eq. (21) satisfies the boundary conditions in Eq. (20). 
Thus we clearly see only left chiral or right chiral fermionic fields
(by assigining $\psi_L (\psi_R)$ to even(odd) fields or vice versa) 
will have zero mode and chiral fermions can thus be identified 
in equivalent 4-D theory.  

But this leads to problem as the boundary points (0, $\pi R$) 
breaks the translational symmetry along the $y$ direction. Thus 
under $S^1/Z_2$ orbifold compactification, the momentum $p_5$ is 
no more conserved and hence the KK number is also not 
conserved. This means that the stability of LKP is no more 
protected by the conservation of KK number.  

However, it can be seen from Eq. (21) that, under a transformation $\pi R$ 
in the $y$ direction, the KK-modes remain invariant when the KK number 
$n$ is even while the KK-modes with $n$ odd change sign. Therefore,
we readily have a quantity, $(-1)^{\rm KK}$ which is a good symmetry 
and is conserved. This is called KK-parity. The conservation of 
this KK-parity ensures that the LKP is stable and therefore is a 
possible candidate for dark matter. In this context, the KK-parity serves the 
same purpose as the $R$-parity in supersymmetric models in terms of 
assuring stability to the dark matter candidate. 

Note that the proposed dark matter candidate $B^1$ (as mentioned before) 
in universal extra dimension model is a bosonic neutral particle whereas 
the candidate (neutralino ($\chi$)) in supersymmetric theory 
is a fermionic neutral particle. This dark matter candidate $B^1$ has been 
explored in several works (see e.g. \cite{servant1, servant2, dm, dm2}).

There are other possible dark matter candidates proposed from other models 
too. One such recently proposed candidate is lightest 
inert particle or LIP from the so called `Inert Doublet' model \cite{lip}.
This LIP dark matter has also been explored (see e.g. \cite{dm3}).
  
\section{Detection of Dark Matter}

As the dark matter has no or very minimal interaction, it is extremely
difficult to detect them. There are two types of detection processes, 
namely direct detection and indirect detection. In direct detection, 
the scattering of dark matter off the nucleus of the detecting material 
is utilised. As this cross-section is very small, the energy deposited 
by a dark matter candidate on the detector nucleus is also very small.
In order to measure this small recoil energy ($\sim$ keV or less) 
of the nucleus, a very low threshold detector condition is required.
In the indirect detection, the annihilation product of dark matter is 
detected. If the dark matter is entrapped by the solar gravitational field,
they may annihilate with each other to produce a standard model particle
such as neutrino. Such neutrino 
signal, if detected, is the signature of dark matter in the indirect 
process of their detection. In what follows, we will discuss the 
direct detection.

Differential detection rate of dark matter per unit detector mass can be
written as
\begin{equation}
\frac {dR} {d|{\bf q}|^2} = N_T \Phi \frac {d\sigma} {d|{\bf q}|^2} \int f(v) dv\end{equation}
where $N_T$ denotes the number of target nuclei per unit mass of the detector,
$\Phi$ - the dark matter flux, $v$ - the dark matter velocity in the 
reference frame of earth with $f(v)$ - its distribution. The integration
is over all possible kinematic configurations in the scattering process.
In the above, $|\bf q|$ is the momentum transferred to the nucleus in
dark matter-nucleus scattering. Nuclear recoil energy $E_R$ is
                                                                                
\begin{eqnarray}
E_R &=& |{\bf q}|^2/2m_{\rm nuc} \nonumber \\
    &=& m^2_{\rm red} v^2 (1 - \cos\theta)/m_{\rm nuc}  \\
m_{\rm red} &=& \frac {m_\chi m_{\rm nuc}} {m_\chi + m_{\rm nuc}}
\end{eqnarray}
where $\theta$ is the scattering angle in dark matter-nucleus centre of 
momentum frame, $m_{\rm nuc}$ is the nuclear mass and $m_\chi$ is the 
mass of the dark matter.
                                                                                
Now expressing $\Phi$ in terms of local dark matter density $\rho_\chi$,  
velocity $v$ and mass $m_\chi$ and writing $|{\bf q}|^2$ in terms
of nuclear recoil energy $E_R$ with noting that $N_T = 1/m_{\rm nuc}$,
Eq. (22) takes the form
                                                                                
\begin{eqnarray}
\frac {dR} {dE_R} &=& 2 \frac {\rho_\chi} {m_\chi} \frac {d\sigma}
{d |{\bf q}|^2} \int_{v_{min}}^\infty v f(v) dv, \nonumber \\
v_{\rm min} &=& \left [ \frac {m_{\rm nuc} E_R} {2m^2_{\rm red}} \right ]^{1/2}
\end{eqnarray}
                                                                                
Following Ref. \cite{jungman} the dark matter-nucleus differential 
cross-section for the scalar interaction can be written as
\begin{equation}
\frac {d\sigma} {d |{\bf q}|^2} = \frac {\sigma_{\rm scalar}}
{4 m_{\rm red}^2 v^2} F^2 (E_R) \,\,\, .
\end{equation}
In the above $\sigma_{\rm scalar}$ is dark matter-nucleus scalar cross-section
and $F(E_R)$ is nuclear form factor given by \cite{helm,engel}
\begin{eqnarray}
F(E_R) &=& \left [ \frac {3 j_1(qR_1)} {q R_1} \right ] {\rm exp} \l ( \frac {q^2s^2}
{2} \r ) \\
R_1 &=& (r^2 - 5s^2)^{1/2} \nonumber \\
r &=& 1.2 A^{1/3} \nonumber
\end{eqnarray}
where thickness parameter of the nuclear surface is given by $s \simeq 1$ fm,
$A$ is the mass number of the nucleus and $j_1(qR_1)$ is the spherical
Bessel function of index 1.
                                                                                
The distribution $f(v_{\rm gal})$ of dark matter velocity $v_{\rm gal}$
with respect to
galactic rest frame, is considered to be of Maxwellian form. The
velocity $v$ (and $f(v)$) with respect to earth rest frame can then be obtained
by making the transformation
\begin{equation}
{\bf v} = {\bf v}_{\rm gal} - {\bf v}_\oplus
\end{equation}
where $v_\oplus$ is the velocity of earth with respect to galactic rest
frame and is given by
\begin{eqnarray}
v_\oplus &=& v_\odot + v_{\rm orb} \cos\gamma \cos \l (\frac {2\pi (t - t_0)}
{T} \r )
\end{eqnarray}
In Eq. (29), $T = 1$ year, the time period of earth motion around the sun,
$t_0 \equiv 2^{\rm nd}$ June, $v_{\rm orb}$ is earth orbital speed and
$\gamma \simeq 60^o$ is the angle subtended by earth orbital
plane at galactic plane. The speed of solar system $v_\odot$ in the
galactic rest frame is given by,
\begin{eqnarray}
v_\odot &=& v_0 + v_{\rm pec}
\end{eqnarray}
where $v_0$ is the circular velocity of the Local System at the position of
Solar System and $v_{\rm pec}$ is speed of Solar System with respect to
the Local System. The latter is also called peculiar velocity and its value
is 12 km/sec. The physical range of $v_0$ is given by
\cite{pec1,pec2}
$170\,\, {\rm km/sec} \leq v_0 \leq 270$ km/sec (90 \% C.L.).
Eq. (29) gives rise to annual modulation of dark matter signal reported
by DAMA/NaI experiment \cite{dama}. This phenomenon of annual 
modulation can be elaborated a little more.
Due to the earth's motion around the sun, the directionality of the
earth's motion changes over the year. This in turn induces an annual
variation of the WIMP dark matter speed relative to the earth (maximum when the
earth's rotational velocity adds up to the velocity of the Solar
System and minimum when these velocities are in opposite directions).
This phenomenon imparts an annual variation of dark matter detection rates
at terrestrial detectors. Therefore investigation of annual variation
of WIMP detection rate is a useful method to confirm the WIMP dark matter 
detection.

Defining a dimensionless quantity $T(E_R)$ as,
\begin{equation}
T(E_R) = \frac {\sqrt {\pi}} {2} v_0 \int_{v_{\rm min}}^\infty \frac {f(v)}
{v} dv\,\,
\end{equation}
and noting that $T(E_R)$ can be expressed as \cite{jungman}
                                                                                
\begin{equation}
T(E_R) = \frac {\sqrt {\pi}} {4v_\oplus} v_0 \l [ {\rm erf} \l ( \frac
{v_{\rm min} + v_\oplus} {v_0} \r ) -  {\rm erf} \l ( \frac
{v_{\rm min} - v_\oplus} {v_0} \r ) \r ]
\end{equation}
we obtain from Eqs. (25) and (26)
\begin{eqnarray}
\frac {dR} {dE_R} &=& \frac {\sigma_{\rm scalar}\rho_\chi} {4v_\oplus m_\chi
m_{\rm red}^2} F^2 (E_R) \l [ {\rm erf} \l ( \frac
{v_{\rm min} + v_\oplus} {v_0} \r ) \r. \nonumber \\
&&\l. - {\rm erf} \l ( \frac
{v_{\rm min} - v_\oplus} {v_0} \r ) \r ]
\end{eqnarray}
The total local dark matter
density $\rho_\chi$ is generally taken to be 0.3 GeV/cm$^3$.
The above expression for differential rate
is for a monoatomic detector like Ge but it can be easily extended for
a diatomic detector like NaI as well.
                                                                                
The measured response of the detector by the scattering of dark matter 
off detector
nucleus is in fact a fraction of the actual recoil energy. Thus, the actual
recoil energy $E_R$ is quenched by a factor $qn_X$ (different for different
nucleus $X$) and we should express differential rate in Eq. (33) in terms of
$E = qn_XE_R$. 


Thus the differential rate in terms of the observed recoil energy 
$E$ for a monoatomic detector like Ge detector 
can be expressed as
\begin{equation}
\frac {\Delta R} {\Delta E} (E) =
\dis\int^{(E + \Delta E)/qn_{\rm Ge}}_{E/qn_{\rm Ge}}
\frac {dR_{\rm Ge}} {dE_R} (E_R) \frac {dE_R} {\Delta E}
\end{equation}
and for a diatomic detector like NaI, the above expression takes the form
\begin{eqnarray}
\frac {\Delta R} {\Delta E} (E) &=&
a_{\rm Na} \dis\int^{(E + \Delta E)/qn_{\rm Na}}_{E/qn_{\rm Na}}
\frac {dR_{\rm Na}} {dE_R} (E_R) \frac {dE_R} {\Delta E}  \nonumber \\
&+&a_{\rm I} \dis\int^{(E + \Delta E)/qn_{\rm I}}_{E/qn_{\rm I}}
\frac {dR_{\rm I}} {dE_R} (E_R) \frac {dE_R} {\Delta E}
\end{eqnarray}
where $a_{\rm Na}$ and  $a_{\rm I}$ are the mass fractions of Na and I
respectively in a NaI detector. 
$$
a_{\rm Na} = \frac {m_{\rm Na}}
{m_{\rm Na} + m_{\rm I}} = 0.153 \,\,\,\,\,
a_{\rm I} = \frac {m_{\rm I}}
{m_{\rm Na} + m_{\rm I}} = 0.847
$$

The differential detection rates $\Delta R/\Delta E$ (/kg/day/keV) 
can thus be calculated for the case of a particular detector 
material.
 

There are certain ongoing experiments and proposed experiments 
for WIMP direct search. The target materials generally used are 
NaI, Ge, Si, Xe etc. NaI (100 kg) is used for DAMA experiment
and near future LIBRA (Large sodium Iodine Bulk for RAre processes) 
experiment (250 kg of NaI) \cite{dama}. These set ups are at 
Gran Sasso tunnel in Italy. 
The DAMA collaboration claimed to have detected this 
annual modulation of WIMP through their direct WIMP detection experiments.
Their analysis suggests possible presence of dark matter with mass 
around 50 GeV. This result is far below the range of LKP mass. 
The Cryogenic Dark Matter Search or CDMS detector employs low temperature 
Ge and Si as detector materials to detect WIMP's via their elastic
scattering off these nuclei \cite{cdms}. This is housed in a 10.6 m 
tunnel ($\sim 16$ m.w.e) at Stanford Underground Facility 
beneath the University of Stanford. 
Although their direct search results are compatible with  3-$\sigma$
allowed regions for DAMA analysis, it excludes DAMA results if standard 
WIMP interaction and a standard dark matter halo is assumed. 
CDMS II experiment \cite{cdms2}
is located at the Soudan underground laboratory at a depth of 780 metres 
(2090 metre water equivalent). The 
EDELWEISS dark matter search experiment which also uses cryogenic
Ge detector at Frejus tunnel, 4800 m.w.e under French-Italian Alps 
observed no nuclear recoils in the fiducial volume \cite{edelweiss}. 
This experiment excludes DAMA results at more than 99.8\% C.L. 
The lower bound of recoil energy in this experiment was 20 keV. 
The second stage of EDELWEISS experiment is EDELWEISS II \cite{edelweiss2}
where a higher detection mass is to be used with low radioactive 
background.
The Heidelberg Dark Matter Search (HDMS) uses in their inner detector,
highly pure $^{73}$Ge crystals \cite{klapdor0} and with a very 
low energy threshold. They have made available their
26.5 kg day analysis. The recent low threshold experiment GENIUS
(GErmenium in liquid NItrogen Underground Setup) \cite{klapdor1} 
at Gran Sasso tunnel in Italy
has started its operation. Although a project for $\beta\beta$-decay search, 
due to its very low threshold (and expected to be reduced futher) 
GENIUS is a potential detector for WIMP direct detection 
experiments and for detection 
of low energy solar neutrinos like pp-neutrinos or $^7$Be neutrinos. 
In GENIUS experiment highly pure $^{76}$Ge is used as detector material.
For dark matter search, 100 kg. of the detector material is suspended 
in a tank of liquid nitrogen. The threshold for Germenium detectors 
is around 11 keV. But for GENIUS, this threshold will be reduced to 
500 eV. The proposed 
XENON detector \cite{xenon} consists of 1000 kg of $^{131}$Xe 
with 4 keV threshold. 

\section{Discussions}

The possible nature of the still unknown and overwhelming dark matter  
is discussed. Different theories predict different possibilities 
of dark matter candidates. Due to space constraints, the calculation of 
relic densities of such candidates could not be addressed. The theoretical 
calculation for direct detection rates in case of a detector material is
also outlined. The experimental
detection, if conclusively confirmed, will not only help us understand
the nature and the particle constituents of dark matter, also it will open new 
vistas in understanding the fundamental laws of nature.

\end{document}